\documentclass[aps,prl,showpacs,notitlepage,reprint]{revtex4-1}
\usepackage{graphicx}% Include figure files
\usepackage{dcolumn}% Align table columns on decimal point
\usepackage{amsmath}
\usepackage{bm}% bold math
\usepackage{todonotes}
\usepackage[utf8]{inputenc}
\usepackage{amsmath}
\usepackage{natbib}
\usepackage{graphicx}
\usepackage{caption}
\usepackage{subcaption}
\usepackage{lipsum}
\usepackage{float}

\newcommand{\beq}{\begin{equation}}
\newcommand{\eeq}{\end {equation}}
\newcommand{\bea}{\begin{eqnarray}}
\newcommand{\eea}{\end{eqnarray}}

\begin{document}
\title{Reply to Comment by K. Forbes on `The super-chirality of vector twisted light' by M. Babiker, J.Yuan, K. Koksal and V. E. Lembessis; Optics Communications 554, 130185 (2024)}
\author{{\rm {M. Babiker}}$^{1,*}$, J. Yuan$^{1}$, K. Koksal$^{2}$, V. E. Lembessis$^{3}$}
\affiliation{$^1$School of Physics Engineering and Technology, University of York, YO10 5DD, UK}
\affiliation{$^2$Physics Department, Bitlis Eren University, Bitlis, Turkey}
\affiliation{$^3$Quantum Technology Group, Department of Physics and Astronomy, College of Science, King Saud University, Riyadh 11451, Saudi Arabia}
\vspace{10mm}
%\affiliation{$^*$ Corresponding author: m.babiker@york.ac.uk}
\vspace{10mm}
\vspace{10mm} 
\date{\today}
\begin{abstract}
We respond to the recent comment in Optics Communications by Kayn Forbes on our recent Optics Communications article and we  maintain that, contrary to what Forbes claims, substantial superchirality exists as a property of the $m\geq 1$ higher order Poincare modes.  Forbes arguments are based on misconceptions and analytical errors, leading to erroneous results and unjustified criticism.
\end{abstract}
\maketitle
The recent comment by Kayn Forbes, Ref.\cite{forbes2024} (to be referred to as KF) on our optics communications Ref.\cite{babiker2024} is focused on the following criticism points: 
\begin{enumerate}
\item{ that there is a difference in the sign of our calculated helicity density compared to KF's helicity density;}
\item{ that the helicity density expression evaluated by KF does not have the same form as our expression;}
\item{ that no superchirality advantage can be offered by higher order Poincare modes;}
\item{ that in strong focusing there is an upper limit that can be realised for the topological order, namely $m<2$.}
\end{enumerate}
Here we respond to each point in turn and conclude that KF's criticism is unjustified.

As to point (1), the minus sign stems from different convention we adopted as regards left-hand and right-hand circular polarisation. It is well known that there are two opposing historical conventions on this. KF followed the convention that for right-hand circular polarisation the ratio $E_y/E_x=-i$ and for left-hand circular polarisation this ratio is $+i$ \cite{born1999}. Our convention follows the equally prevalent convention which is the opposite of this, as in Ref.\cite{saleh1991} (see page 198, TABLE 6.1-1). We agree that this simply leads to an overall minus sign, so we consider the comment by KF on this as inconsequential.

As to point (2), we maintain that our expression for the helicity density is the correct one, following extensive checking by us of the formalism. The expression of KF's helicity density can be re-arranged to read as follows:
\begin{widetext}
\beq
 {\bar {\eta}}({\bf r})=-\frac{\epsilon_0 \omega}{4kk_z}\cos{(\Theta_P)}\left\{k_z^2|{\cal F}|^2+|{{\cal F}}'|^2
 +m^2\frac{|{{\cal F}}|^2}{\rho^2}+2m\frac{{{\cal F}}'{{\cal F}}}{\rho}\right\},
 \label{kayn}
\eeq
\end{widetext}
This emerged from the evaluation by KF of the helicity density using the \underline{incorrect} expression 
\beq
{\bar{\eta}}=-(\epsilon_0\omega/2)\Im[{\bf E}^*\cdot{\bf B}],\label{kaynshel}
\eeq
which coincides with KF's helicity density definition in his recent ArXiv paper \cite{forbes2024b}. Our result for the helicity density is
\begin{widetext}
\beq
 {\bar {\eta}}({\bf r})=\frac{\epsilon_0 c}{4\omega}\cos{(\Theta_P)}\left\{2k_z^2|{\tilde{\cal F}}_{m,p}|^2+|{\tilde{\cal F}}'_{m,p}|^2
 +m^2\frac{|{\tilde{\cal F}}_{m,p}|^2}{\rho^2}+2m\frac{{\tilde{\cal F}}_{m,p}'{\tilde{\cal F}}_{m,p}}{\rho}\right\}
 \label{heldens3}
\eeq
\end{widetext}
which emerged from the standard definition
\beq
{\bar \eta}=-\{\epsilon_0 c/(4\omega)\}\Im[{\bf E}^*\cdot{\bf B}]\label{ours}
\eeq
Besides any insignificant differences in the $\cal F$ functions, namely $\cal F$ by KF in Eq.(\ref{kayn}) and ours ${\tilde{\cal F}}_{m,p}$ in Eq.(\ref{heldens3}), the expression between the brackets as displayed in Eq.(\ref{kayn}) almost coincides with ours in Eq.(\ref{heldens3}) but KF's expression is incorrect by a missing factor of 2 in the first term.  It is also clear that the starting definition Eq.(\ref{kaynshel}) of the helicity density does not have the correct dimensions, which should be `angular momentum per unit volume' (Jsm$^{-3}$). The dimensions of KF's definition of the helicity density, Eq.(\ref{kaynshel}), are Jm$^{-4}$, which is incorrect.

As to point (3), we emphasise that our Fig. 2 is correct apart from the inconsequential difference of sign, as we explained in point (1) above. The plots show the variations of the helicity density with $\rho/w_0$, the radial coordinate in units of $w_0$, for different order $m$. This applies to any point on any order m Poincar\'e unit sphere  along its longitude spanning $(\Theta_P,\Phi_P=0)$ and at each point on the order $m$ longitude it is simply multiplied by the $\cos(\Theta_P)$ factor. The figure  demonstrates that the superchirality density is evident even for $m=1$, where a strong peak exceeding the zero order magnitude is predicted at the core $\rho=0$ and we attribute this to the process of spin-to-orbit conversion for $m=1$.  In the figure, we restricted ourselves to the cases where $k_zw_0<1$, which is valid for a waist parameter consistent with numerical aperture NA $\approx 0.35$ as in \cite{lembessis2024}.

 As to point (4), KF proceeded to put forward arguments based on a statement by Roux \cite{roux2003} that {\it{` the net topological charge in an area cannot exceed the circumference of that area divided by the wavelength'}}.  KF did not offer any sensible justification in this context  and concludes that there is an upper limit of the topological charge, namely $m<2$, when the beam waist is in the sub-wavelength regime. This appears to mean that the circular path accommodates an integer number $m$ of half wavelengths as in a standing wave on a circle, with $m$ interpreted as the topological charge.  Roux's argument is only valid for a uniform helicity density, but LG modes have a non-uniform helicity density.  Furthermore, even when it is averaged to produce an equivalent uniform helicity density, the effective radius, for example for doughnut modes, is not bounded by $w_0$, but by the radius of the doughnut which scales with the topological charge  as $\sqrt{m}$ \cite{padgett1995,curtis2003}. The proposed upper limit contradicts experimental evidence that focused optical vortex modes of higher topological order have been created in the laboratory.  There are reports \footnote{Porfirev, A.P., Ustinov, A.V. and Khonina, S.N. Polarization conversion when focusing vortex beams. Sci Rep 6, 6 (2016). https://doi.org/10.1038/s41598-016-0015-2} which showed that radially and azimuthally-polarised Laguerre-Gaussian modes have been created in the laboratory \cite{neuge2015,bouchard2018,hu2020,yan2022b} and are now available using commercial devices consisting of polarisation converters together with phase plates which add an azimuthal phase dependence. \footnote{https://www.thorlabs.com, http://www.arcoptix.com}.  The modes are thus endowed with the phase function $e^{im\phi}$, with an arbitrary winding number $m$.  KF's argument of an upper limit for the winding number thus goes against experimental evidence \cite{pinnell2020}.

Another argument by KF involving the uncertainty principle is equally unpersuasive and goes against both theoretical and experimental evidence that as the focusing becomes stronger, the longitudinal component becomes comparable to the transverse components and cannot be ignored. 

%(JY:  However, there is also an additional wavefront gradient in the azimuth-direction that scales with $m$ for a given beam width, so can be large under moderate focusing which only controls the . This is consistent with the uncertainty argument used by KF except that he forgot that it is controlled by two independent variables ib 2d space of the tansverse plane.))

 KF's Fig.2 resembles our Fig.2 except for the overall minus sign of the helicity density.  However, it must have escaped KF's notice on looking at the plots that even for $m=1$ there is superchirality, which we define as occurring when the density maximum exceeds that of the zero order.  Furthermore the overall factor $\cos\Theta_P$, which varies continuously between ($1$ and -$1$) can  be chosen to be close  to unity, say $0.99$, close to one of the Poincar\'e sphere poles and still signifies a vector vortex mode, so KF's argument that this indicates a substantial reduction below the circular polarisation case, is a weak argument.

KF did not venture to tackle the ultimate step which we proceeded to take, namely to evaluate the integration of the helicity density over a normal beam cross section to obtain the total helicity per unit length. This is a complex, but analytically rewarding procedure, involving a number of non-trivial integrals.  It leads to our final exact result for the total helicity per unit length ${\cal {\bar C}}_{m}$, which is
\beq
{\cal {\bar C}}_{m}
={\cal L}_0\cos{(\Theta_P)}\left(1+\frac{(m+1){\bar\lambda}^2}{w_0^2}\right)
\label{final}
\eeq
where ${\bar\lambda}=\lambda/(2\pi)$, ${\cal L}_0={\cal P}_T/(k_zc^2)$ is a constant for a fixed power ${\cal P}_T$.  
We have emphasised that our result for the helicity per unit length increases as $m$ increases and exhibits superchirality even for beams waists in excess of a wavelength.  

In conclusion, we reiterate that the recent comment in Optics Communications by Kayn Forbes \cite{forbes2024} on our recent optics communications \cite{babiker2024} is unwarranted.  We  maintain that, contrary to what Forbes claims, substantial superchirality exists as a property of $m\geq 1$ higher order Poincare modes.  Forbes arguments are based on analytical errors and misconceptions leading to erroneous results and unjustified criticism.

\bibliography{bibliography2.bib}
\end{document}